# A Permutation-Based Heuristic for "Buy Low, Sell High"

Yair Neuman,[1] Yochai Cohen[2]

[1] *Cognitive and Brain Sciences & Data Science Center, Ben-Gurion University of the Negev, Be'er Sheva 84105, Israel, yneuman@bgu.ac.il*

[2] *Gilasio Coding, Tel-Aviv 6458701, Israel*

**Abstract**

"Buy low, sell high" is one of the basic rules of thumb used in investment, although it is not considered to be a beneficial strategy. In this paper, we show how the appropriate permutation-based representation (i.e., the epistemic form) of a minute-by-minute trading time-series, alongside the use of a simple decision heuristic (i.e., the epistemic game), may surprisingly result in significant benefits. Using our heuristic for selecting seven stocks, we ran two experiments on the data. The results provide empirical support for the possible benefit of using simple decision models in investment, even in the context of minute-by-minute trading.

**Keywords:** natural cognition; interdisciplinary research; simple models; high-frequency trading; heuristics; ordinal patterns; buy low, sell high

## 1- Main Text

"Take the probability of loss times the amount of possible loss from the probability of gain times the amount of possible gain. That is what we're trying to do. It's imperfect, but that's what it's all about." (Warren Buffett cited in [1])

## 1. Introduction

The current sophistication of mathematical finance invites the question of whether complex models are necessary to achieve significant gain in the financial markets. On the one hand, the "quant revolution" points to the advantage that can be gained from using complex models, at least in some contexts [2]. On the other hand, some successful investors, such as Warren Buffett, promote the idea of simplicity. One may argue that the preference for simplicity, as epitomized in the above quotation, is just a post hoc explanation for an underlying complex model formed unconsciously in Buffett's brain. This argument cannot be dismissed. However, the idea of "bounded rationality" [3], the ecological approach to rationality (e.g., [4, 5]), several papers pointing to the shortcomings of trying to predict future events (e.g., [6]), and the surprisingly superior performance of simple models relative to complex ones (e.g., [7, 8]) suggest that simplicity may sometimes be beneficial. In this context, it may be interesting to study how simple models may support profitable decision-making, even in situations imbued with uncertainty, such as high-frequency (i.e., minute-by-minute) trading.

As a classic paper in cognitive psychology explains, the key to any attempt at "problem-solving," from forecasting to decision-making, is to adequately *represent* (i.e., the "epistemic form") the problem and implement the appropriate *procedure* (i.e., the "epistemic game") for this representation [9]. According to the ecological approach to rationality, these procedures, or epistemic games, are mainly heuristics. In this paper, we draw on these ideas to show that a permutation-based representation (i.e., the epistemic form) of a financial time-series and the use of a simple version of the "buy low, sell high" (BLSH) strategy (i.e., the epistemic game), lacking any predictive aspect, can lead to success in high-frequency trading. It must be emphasized that, in presenting these ideas, we are not attempting to compete with sophisticated optimization engines or algorithmic trading. The cognitive science approach that guides this paper is that a simple model using an appropriate representation (i.e., epistemic form) of a time-series of financial data combined with a simple heuristic (i.e., epistemic game) may produce a surprising benefit over the long run regardless of the uncertainty inherent in the financial time-series.

The paper is structured as follows. Building on the work of Neuman, Cohen, and Tamir [10], we first explain the idea of a permutation-based representation of a time-series and its use in forecasting. Next, we show how to use this idea to apply a very simple version of BLSH. We explain how to use the permutation-based representation plus a simple BLSH heuristic in order to select and then trade several stocks. The next section outlines the results of our tests of the model, where we ran two experiments in which we automatically analyzed seven selected stocks. Finally, in the discussion we elaborate the theoretical aspects of our proposed approach by locating it in the contexts of cognition, ecological rationality, and the potential of simple models to guide natural cognition and behavior "in the wild."

## 2. Permutation-based representation of time-series

The analysis of time-series mostly relies on continuous representation of the variable (e.g., stock price). However, it is possible to represent a time-series in terms of *ordinal patterns*. The idea of analyzing ordinal patterns may be traced back to the seminal work of Bandt and Pompe, where a time-series is converted into a series of ordinal permutation patterns [11]. In this section, we first explain this idea and then point to its cognitive importance as detailed by Neuman, Cohen, and Tamir [10].

Given a one-dimensional time-series $S(t)$ of length $N$, we first partition the series into overlapping blocks of length $D$ (the embedding dimension) using a time delay, $\tau$. Consider the following time-series, which uses $D = 3$ and $\tau = 1$:

$S(t) = \{34, 3, 5, 23, 247, 234, 12, 1, 2, 3\}$

This can be broken down into a sequence of overlapping blocks or vectors:

| **34** | **3** | **5** | 23 | 247 | 234 | 12 | 1 | 2 | 3 |

| 34 | **3** | **5** | **23** | 247 | 234 | 12 | 1 | 2 | 3 |

| 34 | 3 | **5** | **23** | **247** | 234 | 12 | 1 | 2 | 3 |

| 34 | 3 | 5 | **23** | **247** | **234** | 12 | 1 | 2 | 3 |

| 34 | 3 | 5 | 23 | **247** | **234** | **12** | 1 | 2 | 3 |

| 34 | 3 | 5 | 23 | 247 | 234 | 12 | 1 | 2 | 3 |

| 34 | 3 | 5 | 23 | 247 | 234 | 12 | 1 | 2 | 3 |

| 34 | 3 | 5 | 23 | 247 | 234 | 12 | 1 | 2 | 3 |

The elements in each vector are then sorted in ascending order and the vector is mapped onto one of $D!$ permutations (i.e., $\pi_i$), each representing the ordinal pattern of the elements. For $D = 3$, there are six possible permutations:

$\pi_1 = \{0,1,2\}$

$\pi_2 = \{0,2,1\}$

$\pi_3 = \{1,0,2\}$

$\pi_4 = \{1,2,0\}$

$\pi_5 = \{2,0,1\}$

$\pi_6 = \{2,1,0\}$

Next, the first partition in the above time-series – {34, 3, 5} – is mapped onto the permutation pattern $\pi_5 = \{2,0,1\}$; the second partition – {3, 5, 23} – is mapped onto $\pi_1 = \{0,1,2\}$; and so on. This results in a *symbolic sequence of permutations*: $\{\pi_s\}$ $s = 1$ ,,, $n$. The mapping of the above time-series therefore produces a time-series of permutations:

{34, 3, 5, 23, 247, 234, 12, 1, 2, 3} → {2,0,1}, {0,1,2}, {0,1,2}, {0,2,1}, {2,1,0}, {2,1,0}, {2,0,1}, {0,1,2}

The idea of mapping (i.e., representing) a time-series of values onto a time-series of permutations may be highly relevant to prediction in natural environments [10].

An important aspect of time-series of permutations concerns the *constraints* imposed on the *transition* from permutation $\pi_N$ to the next overlapping permutation, $\pi_{N+1}$. For example, in the above time-series of permutations, the first transition is from permutation {2,0,1} to permutation {0,1,2}. While we may naively believe that a transition from each of the six abovementioned permutation types to any of the six permutation types is possible, this belief is wrong: each of the six abovementioned permutation types may move to one of only three permutation types. This *inherent* constraint substantially reduces the potential number of transitions from one permutation type to the next, and hence potentially improves the prediction of the $\pi_{s+1}$ permutation in a *symbolic sequence of permutations*: $\{\pi_s\} s = 1,…,n$.

For $D = 3$ and $\tau = 1$, which are the focus of our study, there are only three legitimate transitions for each permutation. For example, the second partition that we have previously identified – {3, 5, 23} – is mapped onto $\pi_1 = \{0,1,2\}$, where the order of the elements is such that $e_1 < e_2 < e_3$. The following partition/permutation ($\pi_{N+1}$) overlaps with the previous two elements of permutation $\pi_N$ and therefore its first two elements must be ordered such that $e_1 < e_2$ and the only degree of freedom is left to the third element. Among the six possible permutation types for $D = 3$ and $\tau = 1$, there are only three permutation types consistent with this constraint: {0,1,2}, {0,2,1}, and {1,2,0}. What is important to realize is that the constraints imposed on the transition from one permutation to the next significantly reduce the uncertainty associated with the next permutation. A list of legitimate transitions (for $D = 3$, $\tau = 1$) from each permutation type to the next is presented in Table 1.

| Permutation | Legitimate Transition To | | |
|---|---|---|---|
| {0,1,2} | {0,1,2} | {0,2,1} | {1,2,0} |
| {0,2,1} | {1,0,2} | {2,0,1} | {2,1,0} |
| {1,0,2} | {0,1,2} | {0,2,1} | {1,2,0} |
| {1,2,0} | {1,0,2} | {2,0,1} | {2,1,0} |

| {2,0,1} | {0,1,2} | {0,2,1} | {1,2,0} |
|---------|---------|---------|---------|
| {2,1,0} | {1,0,2} | {2,0,1} | {2,1,0} |

**Table 1. A list of legitimate transitions for a given permutation type ($D = 3$, $\tau = 1$).**

## 3. Buy low, sell high with a permutation-based representation

Let us represent a time-series of prices as a series of permutations with length 3. Here we may adopt a BLSH strategy: whenever we observe a decline of the price from $t_n$ to $t_{n+1}$, we buy $N \geq 1$ units of the stock; we then aim to sell it in the next step (i.e., $t_{n+2}$) if we observe an increase. There are three permutations where a price decline is observed from $t_n$ to $t_{n+1}$:

$\pi_3 = \{1,0,2\}$

$\pi_5 = \{2,0,1\}$

$\pi_6 = \{2,1,0\}$

These three permutations are visually presented, from left to right, in Figure 1.

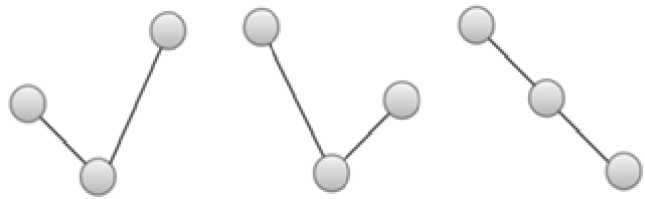

**Figure 1. A visual representation of the three permutations where a decline in price is observed from $t_n$ to $t_{n+1}$.**

In two of these cases (i.e., $\pi_3$ and $\pi_5$), we achieve a profit from selling the stock at the third step and so BLSH is a successful strategy. In one case (i.e., $\pi_6$), we experience a loss when we sell the stock at the third step and so BLSH is a failure. Therefore, the minimal entrance point (MEP) for a rational BLSH strategy as described above is where:

$$100 * \frac{n(\pi_3 + \pi_5)}{n(\pi_3 + \pi_5 + \pi_6)} \geq 51 \tag{1}$$

This means that the relative percentages of $\pi_3$ and $\pi_5$ in the time-series of permutations are equal to or greater than 51. The MEP is a simple heuristic that can be used to select stocks for trading, applying the simple BLSH strategy as described above. To illustrate this idea, we analyzed the time-series of 500 stocks (from September 11, 2017, to February 16, 2018 [12]), represented the time-series as a series of permutations, and computed the MEP of each stock and the 95% confidence interval (CI) for the MEP. After sorting the stocks according to the lower bounds of their MEP 95% CI, we identified the top seven stocks (see Table 2) and used them in our experiments.

| Stock | MEP | SD | Low CI | High CI |
| --- | --- | --- | --- | --- |
| NWS | 87.80 | 5.46 | 86.78 | 88.83 |
| AES | 76.01 | 6.40 | 74.80 | 77.21 |
| NWSA | 74.54 | 6.97 | 73.23 | 75.85 |
| CHK | 73.71 | 5.08 | 72.75 | 74.66 |
| F | 72.14 | 6.22 | 70.97 | 73.31 |
| HPE | 71.56 | 7.43 | 70.16 | 72.96 |
| WU | 70.90 | 6.73 | 69.63 | 72.16 |

**Table 2. The top seven stocks according to the lower bound of the minimal entrance point (MEP). CI = confidence interval; SD = standard deviation.**

According to the idea presented above, selecting stocks according to their MEP should result in a beneficial BLSH strategy. However, the MEP is a necessary but insufficient condition for a successful BLSH strategy. The next phase, which follows Buffett's approach, is to gain a quick estimation of the expected value (i.e., whether the result is likely to be beneficial or not) of applying the BLSH. To compute the expected value, we must somehow estimate the outcome of selling a stock at the third step of each permutation in two scenarios: gain and loss. For simplicity, we can assume that in the context of high-frequency trading, the average delta in price from minute to minute ($\Delta$) is relatively stable and small. Table 3 presents the average $\Delta$ of the abovementioned stocks in absolute percentages.

| Stock | % $\Delta$ |
|---|---|
| NWS | 0.0001 |
| AES | 0.00002 |
| NWSA | 0.00008 |
| CHK | 0.00002 |
| F | 0.00002 |
| HPE | 0.00007 |
| WU | 0.00003 |

**Table 3. The average $\Delta$ of the stock prices.**

We can see that the average $\Delta$ are small. Therefore, for $\pi_5$ (i.e., {2,0,1}), where we observe a decline in the price and then an increase, we may heuristically assume that the increase in the price is half the $\Delta$. Given our ignorance of the size of the decline from rank 2 to rank 0, it is simple to assume that the increase in the price from rank 0 to rank 1 is half the size of the $\Delta$. In this context, buying a stock at the price of $\omega$ and selling at $\omega + 0.5\Delta$ would leave us with a gain of $0.5\Delta$. According to the same logic, for $\pi_3$, the average expected gain is $1.5\Delta$. In all of the cases where we buy a stock after observing a decline in price, the loss expected after observing another decline ($\pi_6$) is $\Delta$. Following a simple expected value (*EV*) analysis:

$$EV = \sum P(Xi) * Xi \tag{2}$$

we should anticipate that if we adopt a simple BLSH strategy, where we gain whenever we encounter $\pi_3$ or $\pi_5$ and lose whenever we encounter $\pi_6$, the expected value will be positive, meaning that applying the BLSH strategy to the series of permutations under the constraint imposed by the MEP is a beneficial strategy. In other words, if the MEP is higher enough and the difference in the stock price is small and consistent enough, then applying the BLSH strategy to the stocks selected using the MEP heuristic should (probabilistically speaking) be beneficial.

In high-frequency trading based on the closing price of a minute-by-minute time-series, there is a basic condition whereby we may calculate the optimal proportion of our total bankroll to use. According to the Kelly criterion (*f\**), the optimal proportion of our total bankroll to use should be:

$$f^* = \frac{bp - q}{b} \tag{3}$$

where *p* is the probability of success, *q* is the probability of failure, and *b* is the odds (or the amount we stand to win relative to the amount we stand to lose). *b* can be estimated by running a simulation on the time-series, but we may also assume that in the short run (i.e., a one-minute difference), the benefits of winning are equivalent to the disbenefits of losing. Given this assumption, we may set *b* to 1. If the opening price of a stock is $11.23 and *p* = 63% (rounded), the Kelly score is 50%. However, as the Kelly score represents the *limit* of a rational bet, a more cautious strategy would be to use a fraction of the Kelly score, such as 0.33 of the *f\**. The proportional Kelly score in this case is 17% and our optimal buy/bet size would be up to 17% of our total bankroll. This limit is important to a rational agent, as in the above context it provides them with a simple heuristic for deciding whether it is beneficial to embark on the BLSH procedure described above. For example, if the price of a single stock were $11.23 and the Kelly score suggested that you should risk up to 17% of your total bankroll (*BR*), then in order to trade with a single stock at a time, your total bankroll would need to be at least:

$$BR = \frac{100*Price}{f^*} = \frac{100*11.23}{17} = \$66 \tag{4}$$

This means that unless you had a bankroll of $66, it would be better not to embark on the BLSH procedure given the above constraints.

The overall heuristic presented so far may be summarized as follows:

1. Search for stocks that satisfy the MEP criterion
2. Rank the stocks according to their MEP scores and select the top $k$
3. For each stock, compute the proportional Kelly score
4. Use the price of a single unit to determine the minimum bankroll
5. If the minimum bankroll is higher than your actual bankroll, then refrain from using the BLSH procedure; otherwise proceed
6. When using the BLSH procedure, buy whenever you observe a decline in price and sell immediately afterward

To test the above heuristic, we analyzed the minute-by-minute stock prices of the seven stocks presented in Table 2. We tested three versions of the heuristic: fight, flight, and freeze. The procedure is detailed in the next section.

## 4. The experiments

### 4.1. Experiment 1

In the first experiment, we used a simple heuristic. For a given selected stock:

1. Define a bankroll, which is the sum that you begin with
2. Start with the first minute of your time-series
3. Your focal point (i.e., the minute you are observing) is always defined as $t_n$
4. If you observe a decrease in price from $t_{n-1}$ to $t_n$ then buy one stock

5. If you observe an increase in price from $t_n$ to $t_{n+1}$ then sell, update your bankroll, and restart the procedure from $t_n+2$ (at this point, $t_{n+2}$ is your focal point)

We experimented with three versions of the heuristic. The strategies differ with regard to what you do when you observe a *decrease* from $t_n$ to $t_{n+1}$ (i.e., when you observe $\pi_6$). The first strategy is **freeze**: whenever you observe a decline in price after you bought the stock, simply wait for a price higher than the one that you observed at the point when you bought, and only then sell. The second strategy is **flight**: whenever you observe another decline in price, sell the stock at $t_{n+1}$, update your bankroll, and restart the procedure (your focal point is now $t_{n+2}$).

The third and most complicated strategy is **fight**, and it is a kind of a rational Martingale strategy. These are ineffective betting strategies where the gambler doubles up their bets upon loses. They are ineffective as the exponential growth of bets leads to bankruptcy. Here we propose a rational version of this strategy. Whenever you observe a decline in price (i.e., $\pi_6$) from the second to the third value of the permutation, you use the proportional Kelly score of your updated bankroll and compute the length of a Martingale buying sequence (e.g., two stocks, four stocks, eight stocks) that you can risk without experiencing bankruptcy. You then start buying according to the Martingale principle. For instance, if you have money then you can buy two stocks at $t_{n+1}$, four stocks at $t_{n+2}$, and so on. You then sell when the price increases or when you run out of the the sum you have originally dedicated to the Martingale, and then restart the procedure. It is important to notice that embarking on this form of rational Martingale strategy is justified only if:

$$P(\pi_6 \to \pi_6) < 0.5 \tag{5}$$

which means that the probability of a transition from $\pi_6$ to $\pi_6$ is lower than 0.5. For example, if we find that

$$P(\pi_6 \to \pi_6) = 0.355 \tag{6}$$

then according to Markov, we may predict that the probability of a second transition to $\pi_6$ is 0.126, that of a third transition to $\pi_6$ is 0.045, that of a fourth transition to $\pi_6$ is 0.016, and so on. This information allows us to approximate the possible length of a series of a continuous decline of price (i.e., a series of $\pi_6$) and the expected length of a Martingale sequence we may be required to apply. Here, we do not delve deeper into this idea as it will be elaborated in another paper.

*4.1.1. Results*

Table 4 presents the prices of the seven stocks at the first point in the time-series, the last point in the time-series, and the maximum point observed throughout the period. For each stock, we identified points where the stock was sold, divided the series into 100 clusters, and presented the bankroll for the end of each of the periods. Figure 2 presents the bankrolls for the three strategies applied to the AES time-series.

| Stock | First | Last | Maximum |
|---|---:|---:|---:|
| NWS | 13.45 | 16.40 | 17.70 |
| AES | 11.24 | 10.47 | 11.95 |
| NWSA | 13.23 | 16.28 | 17.28 |
| CHK | 3.66 | 2.72 | 4.49 |
| F | 11.40 | 10.64 | 13.47 |
| HPE | 13.25 | 16.40 | 17.06 |
| WU | 18.79 | 20.27 | 22.17 |

**Table 4. First, last, and maximum prices ($) of the stocks during the tested period.**

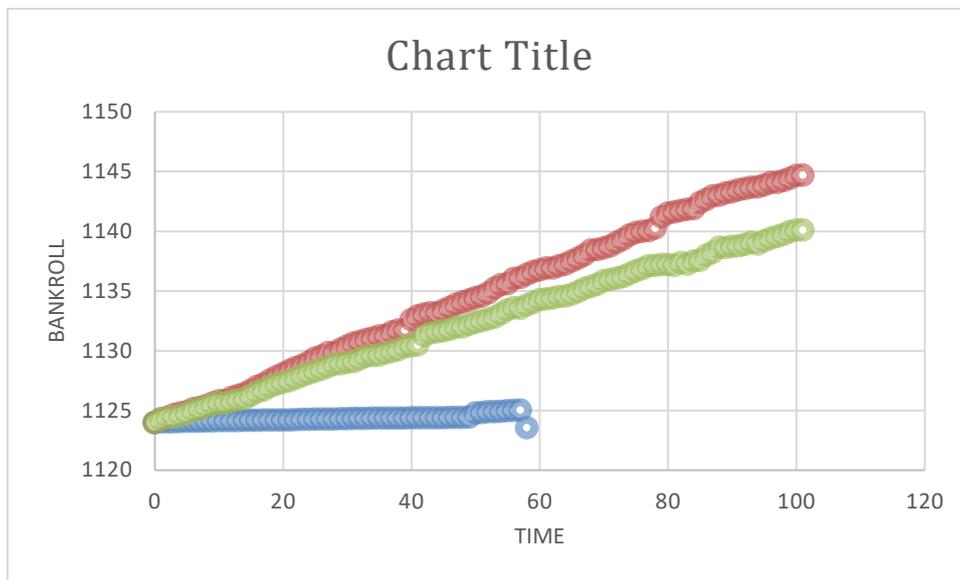

**Figure 2. Bankroll as a function of time for the three versions of the BLSH strategy (Red = Flight, Green = Fight, Blue = Freeze).**

We can see that both the flight and the fight strategies were successful, with the flight strategy outperforming the others. The bankroll for the AES was $1,124. At the end of the trading period, the freeze strategy had lost with a final bankroll of $1,123, while the flight strategy had earned +$21 (1.8%) and the fight strategy had earned +$16 (1.4%). As the freeze version failed in all of the experiments, we will not discuss it further. The AES stock fluctuated in the trading period between a minimum value of $9.89 and a maximum value of $11.94. The price of the stock at the beginning of the trading period was $11.24 and at the end it was $10.47, meaning that buying a stock at the beginning of the period and selling it at the end would have result in a loss. Table 5 presents the gains for the flight strategy (with an average gain of 2.58%) and Table 6 presents the gains for the fight strategy (with an average gain of 2.23%).

| Stock | Initial bankroll | Gain | % gain |
|---|---|---|---|
| NWS | 1,345 | +83.10 | 6.2 |

| Stock | Initial bankroll | Gain | % gain |
|---|---|---|---|
| AES | 1,124 | +21.00 | 1.9 |
| NWSA | 1,323 | +17.55 | 1.3 |
| CHK | 336 | +17.58 | 4.8 |
| F | 1,140 | +19.70 | 1.7 |
| HPE | 1,325 | +17.44 | 1.3 |
| WU | 1,879 | +17.33 | 0.9 |

**Table 5. Percentage gains ($) for the flight strategy.**

| Stock | Initial bankroll | Gain | % gain |
|---|---|---|---|
| NWS | 1,345 | +65.12 | 4.8 |
| AES | 1,124 | +16.00 | 1.4 |
| NWSA | 1,323 | +14.74 | 1.1 |
| CHK | 336 | +17.13 | 4.7 |
| F | 1,140 | +17.52 | 1.5 |
| HPE | 1,325 | +13.79 | 1.0 |
| WU | 1,879 | +15.94 | 0.8 |

**Table 6. Percentage gains ($) for the fight strategy.**

*4.2. Experiment 2*

The above results support the hypothesis that it is beneficial to use a permutation-based representation and to apply the MEP heuristic to this representation. However, in the first experiment, we bought or sold only one unit of the stock at a time. What would have happened if instead of buying a single stock, we had used the relative Kelly score of our bankroll to buy *N* stocks? The second experiment replicated the above procedure with a significant variation: instead of buying a single stock, we bought *N* stocks at a time. We started our trading procedure by using a bankroll equivalent to 100 times the initial price of the stock.

*4.2.1. Results*

Figure 3 presents the trajectories of the bankroll for the NWS stock for the three strategies. The initial bankroll was $1,345 which was equivalent to the value of 100 stocks. Table 7 presents the gains for each stock in the flight strategy and Table 8 presents the gains for each stock in the fight strategy.

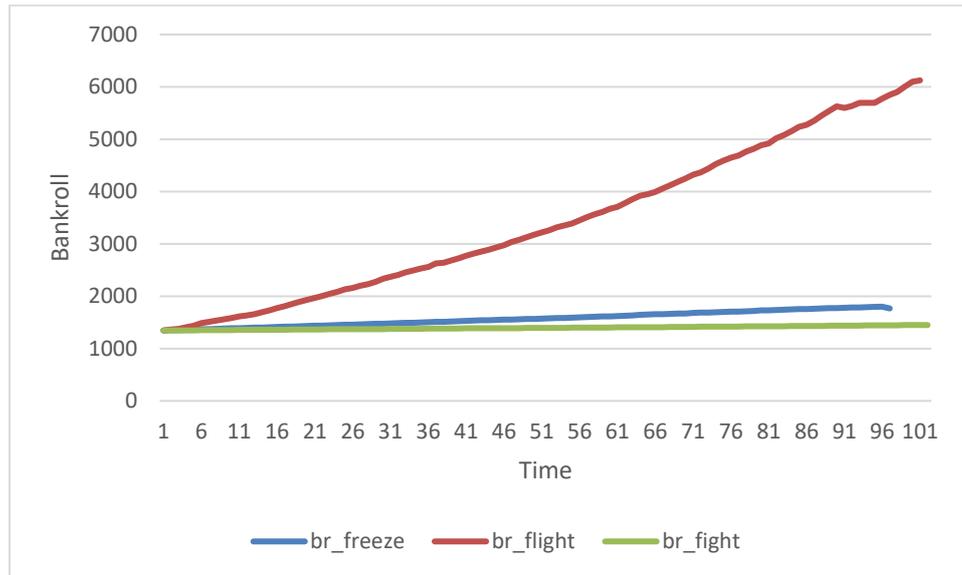

**Figure 3. BR as a function of time (Buying N stocks). (Red = Flight, Green = Fight, Blue = Freeze).**

| Stock | Initial bankroll | Gain | % gain |
|---|---|---|---|
| NWS | 1,345 | +4,778.00 | 355.00 |
| AES | 1,124 | +616.00 | 54.80 |
| NWSA | 1,323 | +391.21 | 29.56 |
| CHK | 336 | +627.92 | 172.00 |
| F | 1,140 | +467.95 | 41.00 |
| HPE | 1,325 | +385.58 | 29.10 |
| WU | 1,879 | +359.78 | 27.00 |

**Table 7. Percentage gains from buying *N* stocks using the flight strategy.**

| Stock | Initial bankroll | Gain | % gain |
|---|---|---|---|
| NWS | 1,345 | +104.00 | 7.70 |
| AES | 1,124 | +34.85 | 3.10 |
| NWSA | 1,323 | +39.80 | 10.87 |
| CHK | 336 | +40.00 | 3.50 |
| F | 1,140 | +26.87 | 2.00 |
| HPE | 1,325 | +33.26 | 2.50 |
| WU | 1,879 | +29.89 | 1.60 |

**Table 8. Percentage gains for buying *N* stocks using the fight strategy.**

We can see that when we trade *N* stocks, the flight strategy outperforms the fight strategy with an average gain of 101.20% vs. 4.47%. This means that at the end of the trading period, our heuristic gained on average 100% profit.

## 5. Discussion

BLSH is one of the basic rules of thumb used in investment [13]. However, the apparent simplicity of this heuristic should be questioned. As argued on a professional investment website, "it's so obvious it sounds like a joke. In reality, it's a lot easier said than done" [14] because predicting the best time to buy or sell stock would involve a "crystal ball" [6] that does not exist for such situations. While it seems obvious to use prediction or forecasting to apply the BLSH strategy, we should recall Makridakis, Hogarth, and Gaba's humorous advice: "By all means, make forecasts – just don't believe them" [15]. The reason that disbelief is advised is that there is no predictive model that allows us to buy at the lowest price and sell at the highest.

In this paper, following a cognitive-heuristic approach, we have shown that representing a time-series of stock prices as a series of permutations may be the first step in a successful heuristic BLSH trading strategy. The use of a permutation-based representation has several benefits, such as moving from absolute to relative values and de-noising the series. Moreover, simply calculating the MEP gives us the first indication as to which of the stocks we should select to trade, and by applying the Kelly criterion we may trade successfully, avoiding the danger of bankruptcy. This approach favors simplicity and avoids the pitfalls of prediction usually associated with BLSH. As explained by Mousavi and Gigerenzer, heuristics are tools that can tackle uncertainty, and they are highly important in situations where the "edge," or the competitive advantage, exists in the uncertainty [16]. Therefore, instead of trying to reduce the uncertainty, the appropriate trading heuristic may use it for good. The uncertainty associated with fluctuations in the prices of stocks may be used by our proposed heuristic to avoid the pitfalls of prediction, and the benefit of its simplicity is apparent. To explain this benefit we may recall the work of Katsikopoulos, Durbach, and Stewart, who argue that "for complex models to outperform simple ones, two conditions must be met: (1) The real-world process must be complex, and (2) the forecaster must be able to model this complexity correctly" [17]. High-frequency trading is complex and we cannot accurately model this complexity. Therefore, avoiding prediction and relying on a simple model and heuristic may be a reasonable approach. This paper therefore presents another instance of where simple epistemic forms (or games) may have a surprising benefit that is worth considering.

## 6. Acknowledgments

The authors would like to thank Yiftach Neuman for helpful elaborations of some formalisms.

## 7. Funding

This research received no funding.

## 8. Author contributions

Conceptualization, Y.N.; methodology, Y.N and Y.C.; software, Y.C.; writing, review, and editing, Y.N. and Y.C. All authors have read and agreed to the published version of the manuscript.

## 9. Conflicts of interest

The author declares that there is no conflict of interest regarding the publication of this manuscript. In addition, ethical principles – including concerning plagiarism, informed consent, misconduct, data fabrication and/or falsification, double publication and/or submission, and redundancies – have been completely observed by the authors.

## 10. Data availability

The time-series of permutations for the tested stocks are available upon request.